\documentclass[reprint,amsmath,amssymb,prl]{revtex4-1}

\usepackage{amsmath}
\usepackage{amssymb}
\usepackage{mathrsfs}
\usepackage{epstopdf}
\usepackage{graphicx}
\usepackage{bm,color,bbm}
\usepackage{hyperref, mathtools}
\usepackage[font = scriptsize]{caption}
\usepackage{subcaption}
\usepackage{subcaption}
\usepackage{float}
\newcommand{\ket}[1]{| #1 \rangle}
\newcommand{\bra}[1]{\langle #1|}
\newcommand{\braket}[2]{\langle #1\vphantom{#2}|
#2\vphantom{#1}\rangle}
\newcommand{\ketbra}[2]{|#1\vphantom{#2}\rangle
\langle#2\vphantom{#1}|}
\newcommand{\matrixel}[3]{\langle #1\vphantom{#2#3}|
#2|#3\vphantom{#1#2}\rangle}

\begin{document}

\title{Compressive Direct Imaging of a Billion-Dimensional Optical Phase-Space}

\author{Samuel H. Knarr}
\email{sknarr@ur.rochester.edu}
\affiliation{Department of Physics and Astronomy, University of Rochester, Rochester, New York 14627, USA}
\author{Daniel J. Lum}
\affiliation{Department of Physics and Astronomy, University of Rochester, Rochester, New York 14627, USA}
\author{James Schneeloch}
\affiliation{Air Force Research Laboratory, Information Directorate, Rome, New York, 13441, USA}
\author{John C. Howell}
\affiliation{Department of Physics and Astronomy, University of Rochester, Rochester, New York 14627, USA}
\affiliation{Racah Institute of Physics, The Hebrew University of Jerusalem, Jerusalem 91904, Givat Ram, Israel}

\date{\today}

\begin{abstract}

Optical phase-spaces represent fields of any spatial coherence \cite{alonso2011wigner,mandel1995optical,waller2012phase}, and are typically measured through phase-retrieval methods involving a computational inversion, interference, or a resolution-limiting lenslet array. Recently, a weak-values technique \cite{lundeen2011direct,lundeen2012procedure,bamber2014observing} demonstrated that a beam's Dirac phase-space \cite{johansen2007quantum,bollen2010direct,hofmann2012complex} is proportional to the measurable complex weak-value, regardless of coherence. These direct measurements require scanning through all position-polarization couplings, limiting their dimensionality to less than 100,000 \cite{bamber2014observing}. We circumvent these limitations using compressive sensing, a numerical protocol that allows us to undersample, yet efficiently measure high-dimensional phase-spaces. We also propose an improved technique that allows us to directly measure phase-spaces with high spatial resolution with scalable frequency resolution. With this method, we are able to easily measure a 1.07-billion-dimensional phase-space. The distributions are numerically propagated to an object in the beam path, with excellent agreement. This protocol has broad implications in signal processing and imaging, including recovery of Fourier amplitudes in any dimension with linear algorithmic solutions \cite{shechtman2015phase} and ultra-high dimensional phase-space imaging.
\end{abstract}

\maketitle


Phase-space representations of light are typically functions of conjugate variables allowing the description of full optical fields of any coherence \cite{alonso2011wigner,mandel1995optical,waller2012phase}. This information has applications in lensless imaging \cite{waller2012phase}, beam shaping \cite{katz2011focusing,mosk2012controlling}, as well as imaging in scattering media \cite{liu20153d}. While measuring a spatially coherent beam's amplitude and phase is a well established field \cite{yamaguchi1997phase}, tomographical measurements of partially coherent beams at high resolution are a laborious challenge, and many require a computational inversion to recover the phase-space distribution. Fortunately, tomography is a standard tool in quantum research used for estimating quantum states \cite{lvovsky2009continuous,james2001measurement}, and the language of quantum mechanics allows us to develop new tools even for classical fields \cite{stoklasa2014wavefront}. 

Recently, a new tomography method was introduced using quantum weak-value techniques \cite{dressel2014colloquium,lundeen2011direct,lundeen2012procedure} to directly measure physical states without optical interference or numerical inversion. Unlike typical quantum tomographical methods that estimate states in terms of the density matrix $\rho$ \cite{james2001measurement} or Wigner function \cite{lvovsky2009continuous,waller2012phase}, the simplest form of the weak-value tomography measures the Dirac phase-space, also known as the Kirkwood-Rihaczek distribution \cite{johansen2007quantum,bollen2010direct}. The Dirac phase-space is a non-Hermitian complex quasi-probability distribution related to the Fourier transform of the density matrix \cite{hofmann2012complex}. When describing quasi-monochromatic and stationary light, optical phase-spaces are functions of four transverse variables: two spatial coordinates and two spatial frequencies. Explicitly, if we have a system with density matrix $\rho$ with transverse positions $\mathbf{x} = (x,y)$ and spatial frequencies $\mathbf{k} = (k_x,k_y)$, the antistandard ordered Dirac representation \cite{bamber2014observing} is
\begin{equation}
S(\mathbf{x},\mathbf{k}) = \textrm{Tr}\left[\ket{\mathbf{k}}\braket{\mathbf{k}}{\mathbf{x}}\bra{\mathbf{x}}\rho\right]= \braket{\mathbf{k}}{\mathbf{x}}\matrixel{\mathbf{x}}{\rho}{\mathbf{k}}
\label{genDiracAS}
\end{equation}
while the standard ordered distribution is
\begin{equation}
S(\mathbf{k},\mathbf{x}) = \textrm{Tr}\left[\ket{\mathbf{x}}\braket{\mathbf{x}}{\mathbf{k}}\bra{\mathbf{k}}\rho\right]= \braket{\mathbf{x}}{\mathbf{k}}\matrixel{\mathbf{k}}{\rho}{\mathbf{x}}
\label{genDiracS}
\end{equation}
where $\mathrm{Tr}[*]$ is the trace. While these distributions are simply complex conjugates, their measurement sequences are different. As with the Wigner function, the marginals, taken by summing over position or frequency variables, give the positive intensity and spectrum of the field, i.e. $\matrixel{\mathbf{x}}{\rho}{\mathbf{x}}$ and $\matrixel{\mathbf{k}}{\rho}{\mathbf{k}}$.

The idea behind these weak-values tomographies is to weakly couple the preselected state of interest (like the transverse state) to another independent degree of freedom, a meter state (such as polarization), and then filter the result by specific measurement outcomes (postselection) in a basis conjugate to the degree of freedom of the preselected state (the Fourier plane of the preselected plane). For example, in the measurement of (\ref{genDiracAS}), analysis of the meter state gives the complex weak-value \cite{lundeen2012procedure}
\begin{equation}
A_{w} = \frac{\braket{\mathbf{k}}{\mathbf{x}}\matrixel{\mathbf{x}}{\rho}{\mathbf{k}}}{\matrixel{\mathbf{k}}{\rho}{\mathbf{k}}}\sim S(\mathbf{x},\mathbf{k})
\end{equation}
with measurable real and imaginary parts. This method allows for direct measurement of the phase-space, removing the need for a computational inversion. Experimental demonstrations of direct measurements of (\ref{genDiracAS}) were first performed measuring transverse pure states \cite{lundeen2011direct} and later mixed (incoherent) states \cite{bamber2014observing}. Weak-value tomographies have also been performed on polarization \cite{salvail2013full,thekkadath2016direct} and orbital angular momentum states \cite{malik2014direct}. 

Direct tomography, while reducing computational complexity, does not necessarily reduce the number of required measurements. Previous demonstrations required scanning through all possible measurements, limiting tests to low-dimensional systems. Here we circumvent previous limitations by incorporating compressive sensing (CS) \cite{candes2008introduction}, which allows us to reduce the resource requirements, especially for high-dimensional states. CS is a numerical method that reconstructs undersampled signals after sampling in a compressive way. CS relies on the assumption of sparsity or approximate sparsity, i.e., the signal of interest has few (or few significant) nonzero components in a predefined basis. 
This assumption works quite well for most signals, as signals of interest usually have structure in some basis that set them apart from random noise. CS has also shown promise in quantum systems \cite{tonolini2014reconstructing,kalev2015quantum} and in the coupling interaction in weak-value tomography \cite{howland2014compressive}.

In this work, we improve upon previous tests by compressively measuring high-dimensional antistandard Dirac spaces. In CS terms, we are using many low-dimensional single pixel cameras to recover the higher dimensional phase-space. We then modify this method for more practical situations to measure the standard Dirac phase-space, allowing for faster acquisition of even larger phase-spaces. With this technique, we easily and efficiently measure phase-spaces of over 1-billion dimensions. Additionally, we expand on previous tests by using a strong polarization-position coupling, which mitigates state-estimation errors \cite{vallone2016strong}.

\begin{figure}[htbp]
  \centering
  \includegraphics[width = .5\textwidth]{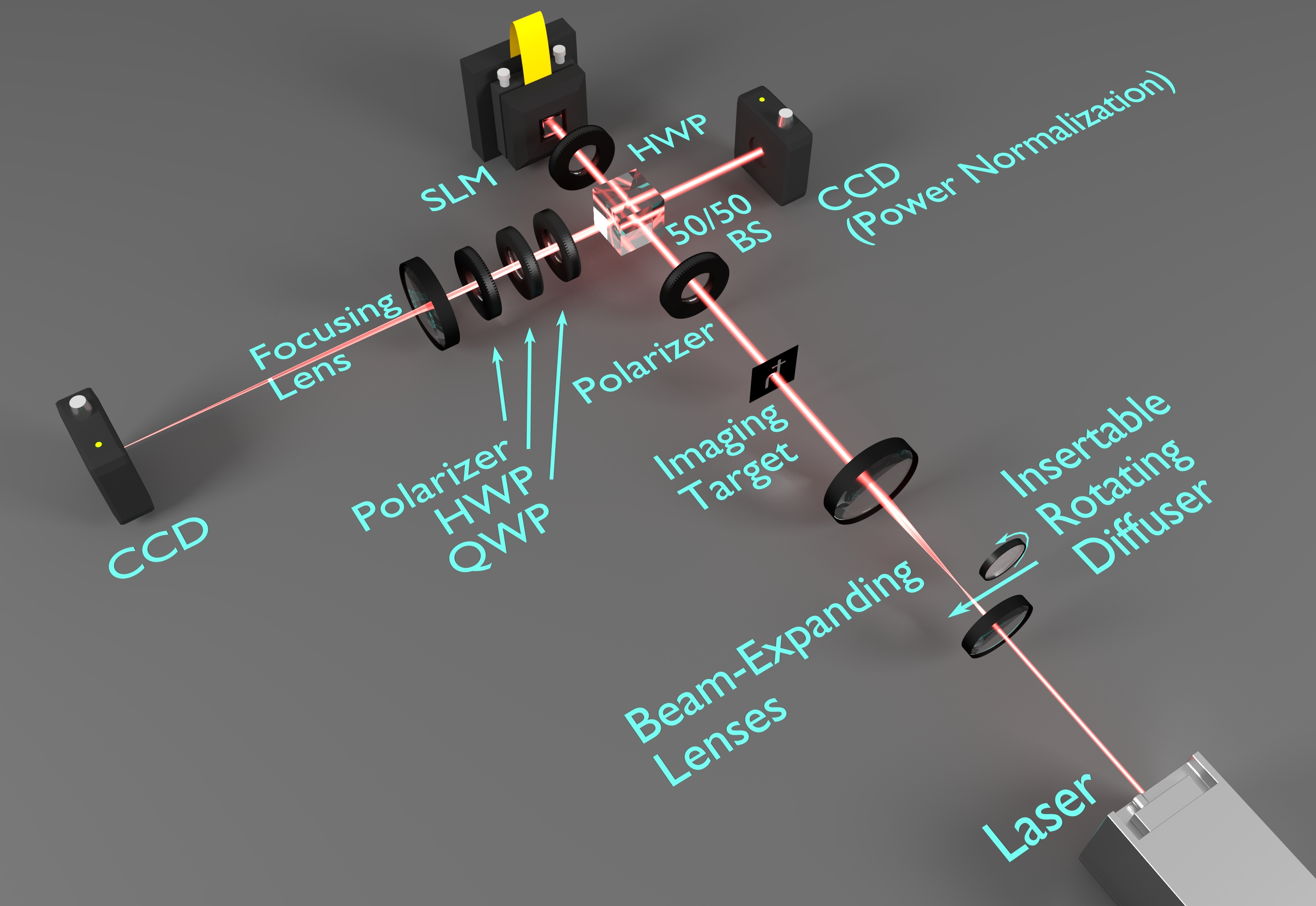}
\caption{\textbf{Experimental Setup:} Light passes through an $\hbar$ paper cutout before propagating 40cm to a horizontal polarizer and SLM. To make the source partially coherent, we insert a rotating glass diffuser in the focal plane of the beam expander. The SLM rotates the polarization at randomly chosen pixels to couple position and polarization. After passing through polarization projection optics, a lens Fourier-transforms the light reflected off the SLM onto a postselection camera. A separate camera acts as a normalizing detector on the reflected port of the beamsplitter in front of the SLM. See Methods for more details on the experiment.}
\label{DiracSetup}
\end{figure}

\begin{figure}[htbp]
  \includegraphics[width = .5\textwidth]{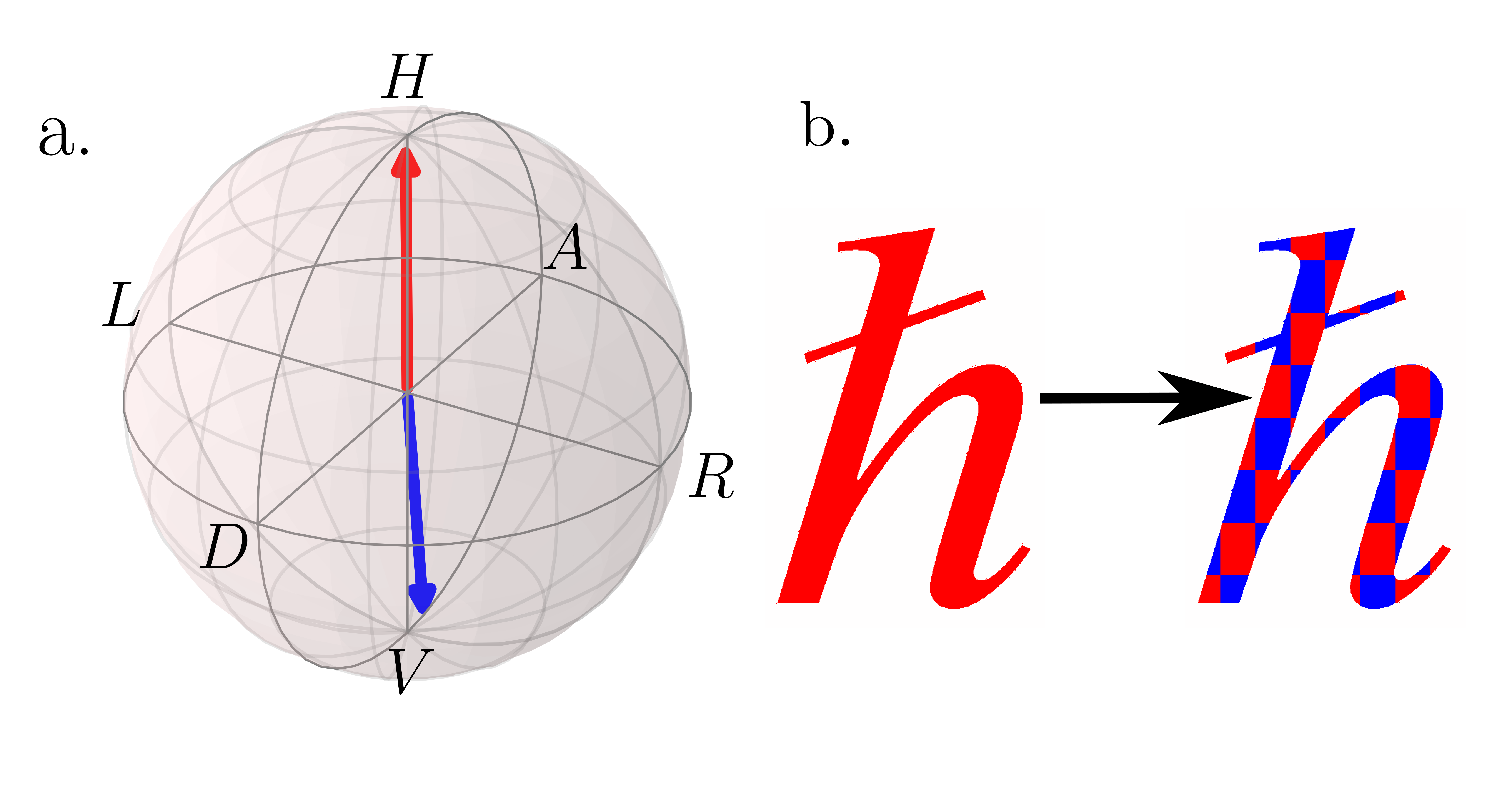}
\caption{\textbf{Polarization Projections:} \textbf{a.} Poincar\'{e} sphere showing the nearly orthogonal polarizations that each pixel occupies as the light leaves our SLM (see Methods for more details on the coupling interaction). \textbf{b.} Here we show an $\hbar$ pixelated into the two outgoing polarizations.}
\label{SphereHbar}
\end{figure}

First we introduce CS into the measurement of (\ref{genDiracAS}). Our experimental setup is shown in Fig. \ref{DiracSetup}. The light passes through an $\hbar$ cutout and propagates freely to a polarizer to prepare the meter state. Since we are interested in the transverse distribution, we couple position to polarization by using a spatial light modulator (SLM) to rotate the polarization by an angle $\theta$ at certain pixels. We apply $M$ masks of $N\times N$ random patterns $f_{i}(\mathbf{x})$, for $i=1,2,...,M$, to the SLM, which couple random positions with polarization. As the measurement is compressive, $M<<N^2$. The SLM's operation is
\begin{equation}
\hat{U}(f_{i}(\mathbf{x}),\theta) = \int\textrm{d}\mathbf{x}\textrm{ }\big[\mathbbm{1}_{\sigma}(1-f_{i}(\mathbf{x}))+f_{i}(\mathbf{x})e^{-i\mathbf{\sigma}\cdot\mathbf{n}\theta}\big]\ket{\mathbf{x}}\bra{\mathbf{x}}
\label{polRot}
\end{equation}
where $\sigma$ is the usual Pauli operator, $\mathbbm{1}_\sigma$ is the Pauli identity, and $\mathbf{n}$ is the axis of rotation. The first term ensures pixels where $f_i(\mathbf{x})=0$ stay horizontally polarized, while the second term rotates the polarization of pixels where $f_i(\mathbf{x})=1$. For the strongest possible coupling, we let $\theta\rightarrow\pi/2$, so these pixels are rotated to vertical polarization. This operation is shown in Fig. \ref{SphereHbar}.


Next, we postselect on different frequency components. This postselection creates an interference between the position and frequency projections of $\rho$ that allows for transverse phase-information retrieval. 
To postselect, a lens Fourier-transforms the light reflected off of the SLM onto a camera, which takes $N\times N$ pictures for each coupling mask on the SLM. Note that the postselection is a pixel-wise operation, meaning that there is a strong frequency projection measurement at each pixel. 
Since the transverse state is coupled to polarization, a polarization analysis (we record an image for each mask of each polarization projection) determines the real and imaginary parts of the transverse Dirac distribution (see Methods for more on the polarization projections). 

If we think of our four-dimensional Dirac phase-space as a 2D space with indexed positions $(x,y)$ as rows and spatial frequencies ($k_x,k_y$) as columns, the combined images of the polarization projection for a single coupling mask correspond to a single row. That is, we measure an $M\times N^2$ compressed phase-space, where each of the $M$ rows corresponds to a coupling mask. Essentially in this 2D reshaping picture, the phase-space is only compressed in one dimension. Each column is then separately reconstructed from the measurement matrix. Thus, reconstruction of the phase-space comes from solving many smaller CS problems.

\begin{figure}[htbp]
  \includegraphics[width = .5\textwidth]{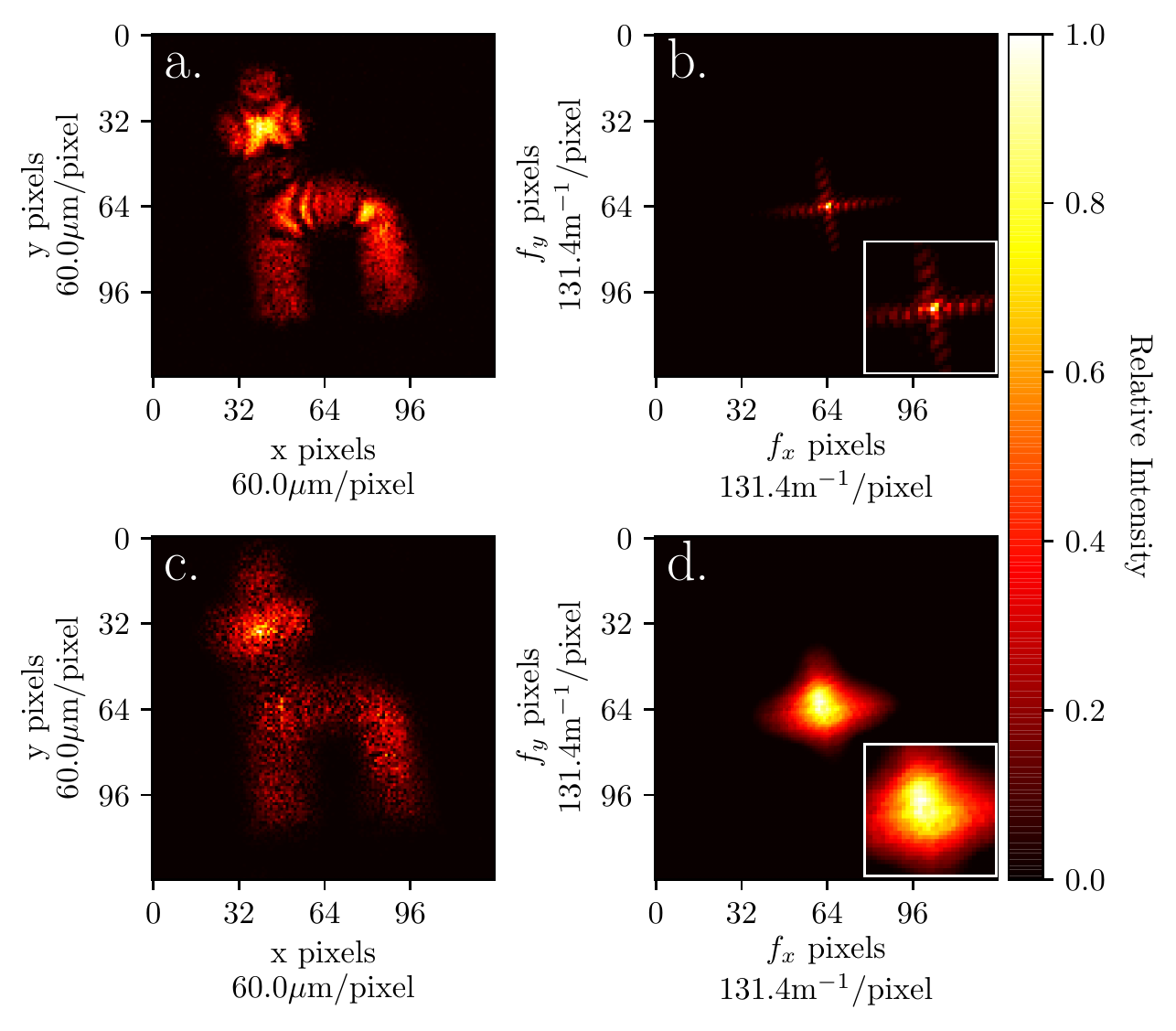}
\caption{\textbf{Marginals of the measured antistandard Dirac distributions:} The top row shows the results from the coherent source; the bottom row contains the marginals from the partially coherent source. All probability distributions have been individually normalized to the same scale for visual clarity. \textbf{a and c.} The position marginals corresponding to the intensities on the SLM. \textbf{b. and d.} The frequency marginals, here in units $\mathbf{k}/2\pi$, showing the diffraction patterns on the camera. The insets show closer views of the center $64\times64$ pixels of the diffraction patterns.}\label{Marginals}
\end{figure}

We use $N=128,$ so that the measured phase-space dimensionality is 268-million (i.e., $128^4$). For our CS solver, we use a total-variation-minimization solver TVAL3 \cite{li2009user}. Unfortunately, this algorithm did not incorporate any of the physical constraints \cite{hofmann2012complex} on the Dirac distributions, which left unphysical artifacts in the reconstructions. The marginals of these distributions are displayed in Fig. \ref{Marginals}; these should be the probability distributions of the light on the SLM and camera. To remove unphysical negative and imaginary values, we show the real part with negative values thresholded to zero (see Methods for more on the data collection and processing). 
As expected after the long free space propagation, we see blurry images of the $\hbar$ and their diffraction patterns. In the coherent source marginals, we can clearly see fringes on the $\hbar$ and a tight diffraction profile. For the partially coherent light, the $\hbar$ is simply blurred and the diffraction pattern is quite broad. 
We perform this measurement using a 20\% sampling rate.
That is, the measurement uses approximately 3300 out of $128^2$ possible coupling projections, recording an image for each pattern and polarization projection.

\begin{figure}[htbp]
  \includegraphics[width = .5\textwidth]{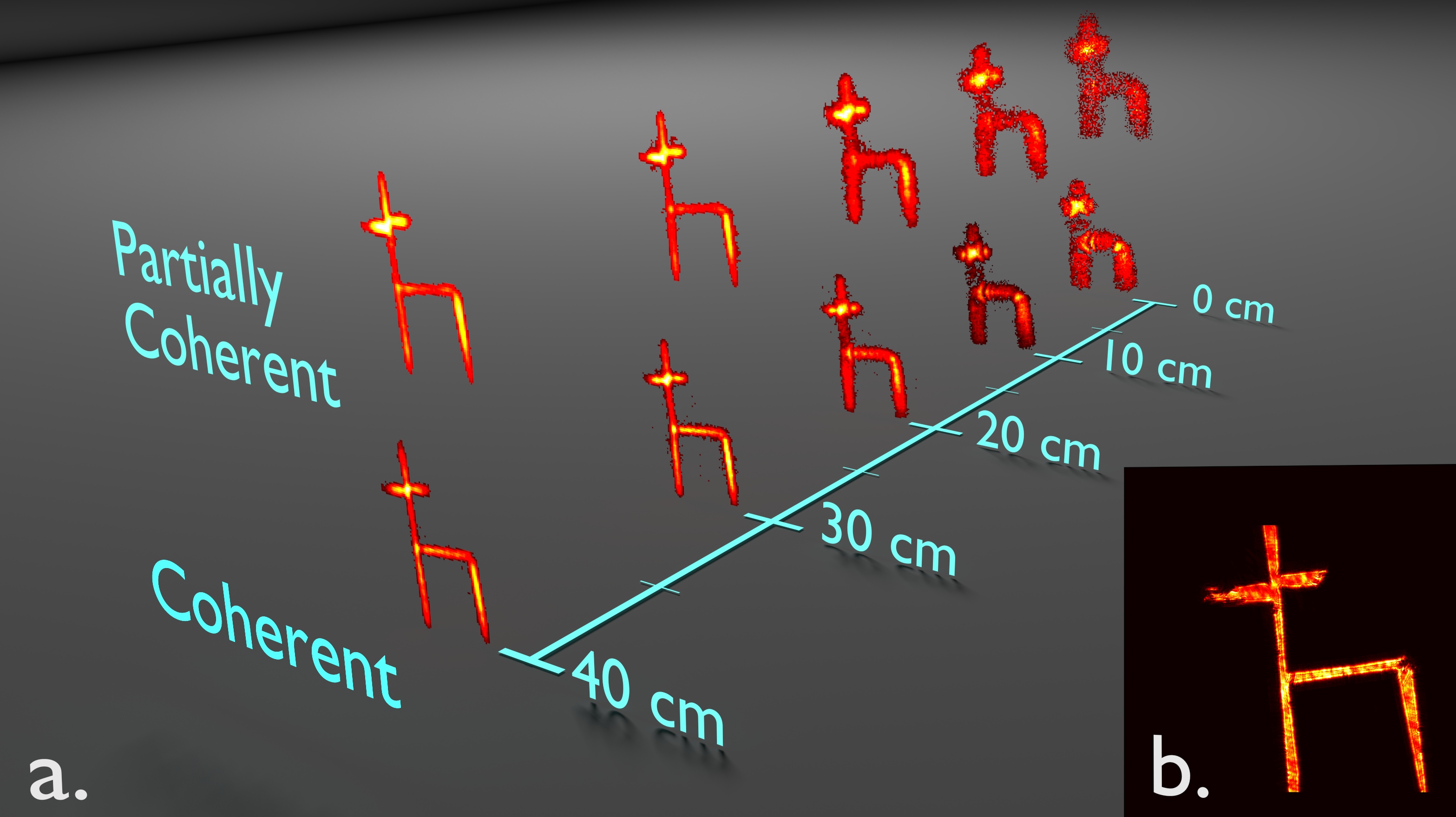}
\caption{\textbf{Comparison of numerical propagation with actual image:} Plots have been normalized to be shown on the same scale. \textbf{a.} The position marginal of the propagated Dirac distributions showing the intensities as functions of distance from the SLM (0cm) to the object (40cm). The upper row shows the partially coherent light, while the lower row shows the coherent light. \textbf{b.} Actual image of the object. The propagation data is in good agreement with the image for both sources.}\label{Propagation}
\end{figure}

To test the reconstruction's accuracy, we numerically propagate the reconstructions back to the $\hbar$ cutout to recover the light at the object. This is done by solving the four-dimensional Bayesian propagation integral \cite{hofmann2012complex}
\begin{equation}
S(\mathbf{x}',\mathbf{k}') = \int \textrm{d}\mathbf{x}\,\textrm{d}\mathbf{k}\, K(\mathbf{x},\mathbf{k};\mathbf{x}',\mathbf{k}') S(\mathbf{x};\mathbf{k})
\label{genProp}
\end{equation}
where $K(\mathbf{x},\mathbf{k};\mathbf{x}',\mathbf{k}')$ is the Dirac phase-space propagator. Since the propagation is through free space, the spatial frequency integrals vanish leaving an integral that can be easily evaluated with standard computational Fourier methods, independent of the spatial coherence of the source (see Methods for a free-space demonstration of Bayesian propagation). After propagating the distributions back to the object plane, we find sharp object images shown in Fig \ref{Propagation}. The position marginals of the propagated distributions are in excellent agreement with an image taken of the cutout. The partially coherent results contain higher levels of background noise, but this is most likely due to the weaker signal after the diffuser. Again we remove unphysical values by showing the real positive part of the marginals.

\begin{figure*}[htbp]
  \includegraphics[width = .8\textwidth]{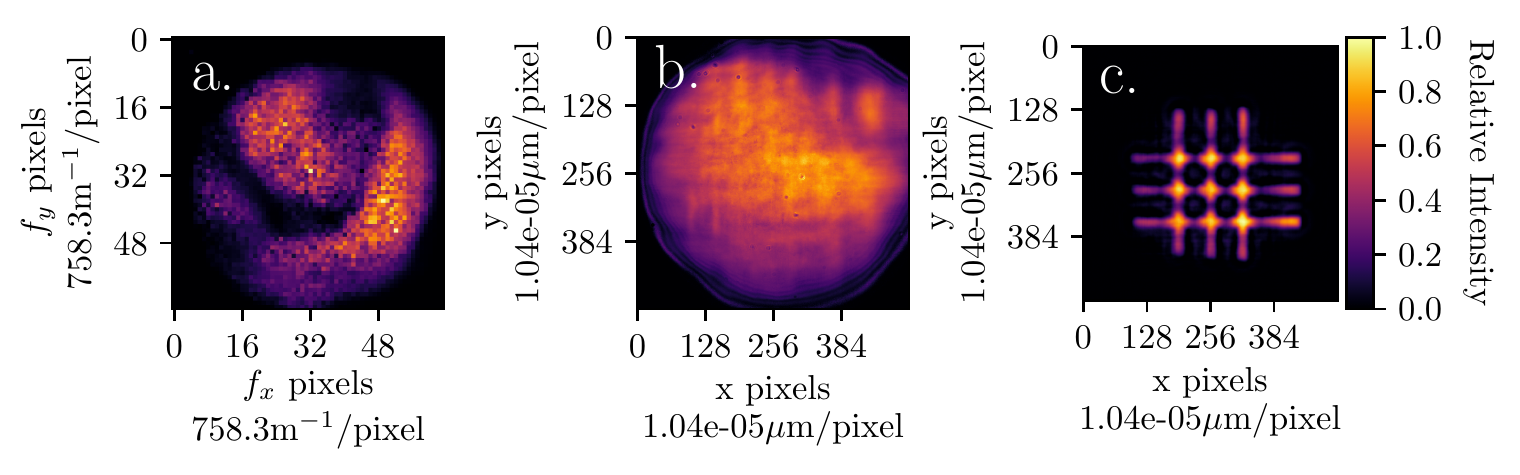}
\caption{\textbf{Measurement of the standard ordered Dirac distribution:} Plots have been normalized to be shown on the same scale. \textbf{a.} The frequency marginal of the Dirac distributions showing the intensities across the SLM. The circular outline is due to an iris blocking the light from the highly reflective edges of the SLM. \textbf{b.} Spatial marginal corresponding to the extremely out-of-focus image on the camera. \textbf{c.} The spatial marginal after propagation, clearly showing the well-defined object.}\label{DiracFT}
\end{figure*}

To modify this experiment to measure (\ref{genDiracS}), we insert a lens after the object to focus the light onto the SLM. This effectively reverses the domains of the previous test; the spatial frequencies are coupled to polarization and the postselection camera captures the transverse positions. All of the previous analysis applies with the appropriate changes. While this is a simple experimental change, it offers several practical advantages. First, we are now able to use the high-resolution postselection camera to gather more spatial dimensions. Since most practical applications do not require such high-resolution spatial frequency information, we can perform faster lower resolution CS scans on the SLM (see Methods for more data collection benefits and processing information). These lower dimensional CS reconstructions also lower the computational burden, while still acquiring a higher dimensional phase-space faster than the previous method.

In our test, we switch our simple cutout for a digital micro-mirror device (DMD) displaying a crossed 3-slit pattern. We displace the DMD 10cm from the focal plane of the lens, severely blurring the image of the object on the camera. We broadly illuminate this object with partially coherent radiation and acquire a 1.07-billion-dimensional phase-space ($512^2$ spatial dimensions and $64^2$ spatial frequencies). Our measured and propagated results are shown in Fig \ref{DiracFT} for a 20\% sample rate. Clearly after propagation we can see a sharp object. 

By using a postselection camera combined with compressive coupling, our required measurements only scale as the resolution of our SLM $N^2$, while measuring an $N^4$ dimensional phase space. Our demonstration here uses relatively simple scenes, and our reconstructions only assume that total variation of the beam across the SLM pixels would be sparse. However, total variation has been shown to work well for natural images, and in practice one would usually know something about the signal of interest allowing them to chose a basis for reconstruction where the signal should have a sparse representation. For example, natural images are known to also have sparse representations in the discrete cosine and wavelet bases. 

We have shown that we can efficiently acquire high-dimensional optical phase-spaces for light of any spatial coherence. With this information, we can numerically propagate the light to any plane for lensless imaging. This goes significantly further than previous demonstrations that only worked in one dimension and required scanning at every coupling pixel \cite{bamber2014observing}. Unlike other phase-space techniques, we did not need to use lenslet arrays or move our detectors, and we were able to directly measure the phase-space elements. Another benefit to this method is that the measured Dirac distribution scales as the product of the number of pixels in the coupling interaction with the number of pixels in the postselection meaning that we can very easily measure extremely high-dimensional phase-spaces. In the future, customized algorithms incorporating physical constraints could give better reconstructions and further reduce the number of measurements and the numerical resources required.

\section{Methods}
\subsection{Antistandard Dirac Experiment}
The coupling interaction uses a Meadowlark Optics XY spatial light modulator (512$\times$512 pixels with a pixel pitch of 15$\mu$m, binned into $128\times128$ pixels). 
We calibrate our system through a polarization tomography \cite{james2001measurement} so that the unrotated state is $\ket{H}$, and the rotated state is as close to orthogonal ($\theta \approx.98\pi$) as possible, such that we can reasonably approximate it as $\ket{V}$. In this way we get the largest coupling between position and polarization. These states are shown on the Poincare sphere in Fig. \ref{SphereHbar}.

Let us examine the measurement interactions in more detail. For simplicity, let the input state be a pure state $\ket{\Psi} = \ket{\psi(\mathbf{x})}\ket{H}$. Applying the operation in Eq. (\ref{polRot}), the input state becomes
\begin{equation}
U(f_{i}(\mathbf{x}),\theta)\ket{\Psi} = \int\textrm{d}\mathbf{x}\psi(\mathbf{x})\left[\left(1-f_{i}(\mathbf{x})\right)\ket{H}+f_{i}(\mathbf{x})\ket{V}\right]\ket{\mathbf{x}}.
\end{equation}
For our patterns, we use randomized Hadamard patterns, which are composed of 1s and -1s. To perform this operation, we split each pattern into a projection with 1s and 0s, and subtract an inverse pattern with the 0s and 1s switched.

We Fourier-transform the light on the face of the SLM using a 250mm lens onto a Thorlabs DCC1545M camera (1280$\times$1024 pixels with 5.2$\mu$m$\times5.2\mu$m pixels we use the center $512\times512$ pixel section binned into $128\times128$ superpixels). Unfortunately, the light quickly saturates the camera since it is in the focal plane. This means that we have to attenuate the light and average several images per patten in order to resolve the high frequency components. For the coherent illumination, we average 16 images taken with 6ms integration times, while for the partially coherent light, we average 64 of these images. The pixels report 8-bit intensity values; to remove background noise, we threshold away any pixel value less than 1-bit. All images are normalized for power fluctuations by using the normalizing camera shown in Fig. \ref{DiracSetup}.

The unnormalized state at each camera pixel is
\begin{equation}
\ket{\psi_{ps}}\sim\tilde{\psi}(\mathbf{k})\ket{H}-\int\textrm{d}\mathbf{x}f_{i}(\mathbf{x})\psi(\mathbf{x})e^{-i\mathbf{k}\cdot\mathbf{x}}(\ket{H}-\ket{V}),
\end{equation}
where $\tilde{\psi}(\mathbf{k})$ is the Fourier transform of $\psi(\mathbf{x})$. For a pure state, the Dirac distribution (\ref{genDiracAS}), can be written as $S(\mathbf{x},\mathbf{k}) = \psi(\mathbf{x})\tilde{\psi}^*(\mathbf{k})e^{-i\mathbf{k}\cdot\mathbf{x}}$ where $\psi^*$ is the complex conjugate of $\psi$. By performing the following polarization projections \cite{vallone2016strong}, we can measure the partially compressed real and imaginary parts of the Dirac distribution at each pixel $(k_x,k_y)$,
\begin{equation}
\begin{aligned}
y_{\mathrm{real},i,\mathbf{k}} &= \int\textrm{d}\mathbf{x}f_{i}(\mathbf{x})\textrm{Re}\{\textrm{S}(\mathbf{x},\mathbf{k})\}
\sim \langle\sigma_x+2\ketbra{V}{V}\rangle_{\psi_{ps}}\\
y_{\mathrm{imag},i,\mathbf{k}} &= -\int\textrm{d}\mathbf{x}f_{i}(\mathbf{x})\textrm{Im}\{\textrm{S}(\mathbf{x},\mathbf{k})\}
\sim \langle\sigma_y\rangle_{\psi_{ps}}.
\label{yvecs}
\end{aligned}
\end{equation}
To more efficiently measure the non-Hermitian operator $\sigma_x+i\sigma_y$, we reduced our number of measurements by decomposing this into a complex sum of Hermitian operators as in \cite{bolduc2016direct}
\begin{equation}
\sigma_x+i\sigma_y = \frac{4}{3}\sum_{q=0}^2 e^{2\pi i q/3} \ketbra{s_q}{s_q}
\end{equation}
where
\begin{equation}
\ket{s_q} = \frac{1}{\sqrt{2}}(\ket{H}+e^{-4\pi i q/3}\ket{V}).
\end{equation}

Equipment limitations added a large amount of overhead time to the measurements. The coherent illumination experiment takes approximately 11 hours, however, the time spent collecting light was only 1.5 hours. The partially coherent test takes 29.5 hours with 5.5 hours of signal collection. Most of this extra time came from the cameras not having internal storage and requiring each image acquisition be transferred to a computer With reasonable additional resources, this large overhead time could be significantly reduced. Simply parallelizing the CS interactions and polarization projections would lead to an eightfold improvement. Additionally, by using faster cameras with onboard storage, we could further reduce our total measurement time for both illumination sources to on the order of minutes or faster.

\subsection{Standard Dirac Experiment}
In this experiment we use the same configuration as above, except with a 250mm focal length lens before SLM such that the SLM is in the focal plane. Our object is a 3-crossed slit pattern broadcast across a DLP Lightcrafter 3000 DMD and is placed 15cm in front of the lens. We flood illuminate the DMD with a red LED, frequency filtered with a 633nm line filter to work with the SLM. An iris was placed just before the SLM to block the light from reflecting off the metal edges, which would only add noise to the signal. The theory and measurement scheme presented above still applies, with simply switching the domains.

This configuration had several practical benefits that shift the required resources from experiment to computation. First since the camera was no longer in the focal plane of the system, we avoid the saturation effects affecting the previous scheme. Thus we do not need to take and average as many images. For this test, we use a 25ms integration time (due to the dimness of the light on the camera), and average only 4 images. Again equipment limitations added to the runtime of the experiment, but the total time spent collecting light was only 12 minutes. With the additional resources discussed above and a more sensitive camera, this measurement could easily be done in seconds.

\subsection{Data Analysis}
In order to utilize fast transforms for efficient CS reconstructions, we use randomized Sylvester-Hadamard patterns. We reconstructed the real and imaginary parts of the Dirac phase-space separately, one component at a time. This limits the number of computations necessary on the high-dimensional distribution. We use the TVAL3 solver \cite{li2009user}, which searches for solutions to the problem
\begin{equation}
\mathrm{arg \,}\mathop{\mathrm{min}}\limits_{\mathbf{x}} \quad \left[ ||\mathbf{D}\mathbf{x}||_1+\frac{\mu}{2}||\mathbf{A}\mathbf{x}-\mathbf{y}||_2^2\right]
\end{equation}
where $\mathbf{D}$ is the discrete gradient across $\mathbf{x}$ and $||*||_p$ is the $L^p-$norm. $\mathbf{A}$ is a matrix containing our projectors, and $\mathbf{y}$ is a vector containing our measurement results. By using this solver, we assume that the total variation of each distribution is sparse across the SLM. We make this assumption because TVAL3 has been shown to work well in image processing because images are often well-defined by their edges and an image's gradient will emphasize edges. It is not unreasonable to guess that it should work in a similar way here. Beyond this, we do not make any assumptions about the sparsity of the Dirac distribution and performed all reconstructions in the pixel basis. 

Since we are reconstructing the Dirac phase-space's real and imaginary part at each camera pixel  separately, theoretically we require $2N^2$ reconstructions. However, when measuring the antistandard ordered Dirac matrix, the measurement matrices $\mathbf{y}$ are quite sparse as many of the spatial frequencies are zeros, so no reconstruction is needed for these components since there is no signal. This is confirmed in Fig. \ref{Marginals}b and \ref{Marginals}d. In total, we only have to perform 4544 reconstructions for the coherent illumination and 2871 reconstructions for the partially coherent light of $128\times128$ pixel images. We would normally expect that the partially coherent light requires more reconstructions than the coherent light since its measurement vector should have more frequency components. However, it is likely that the weaker high-frequency components are too dim for the camera to see above the background noise. This is not quite the case in measuring the standard ordered Dirac distribution, which requires more than 200K reconstructions for each the real and imaginary due to the broad image distribution on the camera. However, these are smaller $64\times64$ reconstructions which can be done quite quickly and in parallel on modern computers.
Next, we performed a Bayesian shrinkage denoising algorithm with soft thresholding \cite{chang2000adaptive} followed by a low (1\%) hard thresholding to remove low-level noise from the Dirac distribution. We then further correct the phase-space elements using a least squares fitting algorithm on each reconstruction result which assumed Gaussian noise in signal \cite{fletcher2013practical}. 

The TVAL3 solver does not incorporate any physical constraints\cite{hofmann2012complex} on the reconstructions, and so experimental noise left unphysical artifacts in the phase-space. The denoising and least squares steps fix some of the values, but the result is still slightly unphysical. This results in marginals having complex and negative values when they should be entirely positive since they are probabilities. From a practical point of view, this does not strongly affect the propagation and refocusing of the light, and by viewing the real positive part of the marginals, we had good agreement in the data shown in Fig \ref{Propagation} and Fig \ref{DiracFT}. In future work, algorithms could be created to handle this problem more physically. 

\subsection{Propagation}
As noted in \cite{lundeen2012procedure,hofmann2012complex}, the Dirac distribution can be propagated in a Bayesian manner. Therefore, using the definition of the Dirac distribution Eq.(\ref{genDiracAS}), for $S(\mathbf{x},\mathbf{k})$ we can find the Dirac representation in the plane $(x',y'),  S(\mathbf{x'},\mathbf{k'})$, through
\begin{equation}
\begin{aligned}
S(\mathbf{x'},\mathbf{k'}) =&\int \textrm{d}\mathbf{x}\,\textrm{d}\mathbf{k}\,\braket{\mathbf{k'}}{\mathbf{x'}}\braket{\mathbf{x'}}{\mathbf{x}}\matrixel{\mathbf{x}}{\rho}{\mathbf{k}}\braket{\mathbf{k}}{\mathbf{k'}} \\ 
=&\int \textrm{d}\mathbf{x}\,\textrm{d}\mathbf{k}\,\frac{\braket{\mathbf{k'}}{\mathbf{x'}}\braket{\mathbf{x'}}{\mathbf{x}}\braket{\mathbf{k}}{\mathbf{k'}}}{\braket{\mathbf{k}}{\mathbf{x}}} S(\mathbf{x};\mathbf{k})
\end{aligned}
\end{equation}
where the propagator $K(\mathbf{x},\mathbf{k};\mathbf{x'}\mathbf{k'})$ of Eq. (\ref{genProp}) is immediately identified. For free space propagation over a distance $z$, the terms in $K$ simplify to
\begin{equation}
\begin{aligned}
&\frac{\braket{\mathbf{k'}}{\mathbf{x'}}}{\braket{\mathbf{k}}{\mathbf{x}}} = e^{i(\mathbf{k}\cdot\mathbf{x}-\mathbf{k'}\cdot\mathbf{x'})}\\
&\braket{\mathbf{k}}{\mathbf{k'}}  = e^{-iz\sqrt{k^2-k_x^2-k_y^2}}\delta(\mathbf{k}-\mathbf{k'})\\
&\braket{\mathbf{x'}}{\mathbf{x}} = \frac{1}{(2\pi)^2}\int \textrm{d}\mathbf{k''} e^{i\mathbf{k''}\cdot(\mathbf{x}-\mathbf{x'})}e^{iz\sqrt{k^2-k_x^{''2}-k_y^{''2}}} 
\end{aligned}
\end{equation}
where the middle term shows that through free-space propagation, momentum is conserved. Putting this altogether and rearranging terms results in the propagation equation
\begin{equation}
\begin{aligned}
S(\mathbf{x},\mathbf{k}) &= \frac{1}{(2\pi)^2}e^{-iz\sqrt{k^2-k_x^2-k_y^2}}e^{-i\mathbf{k}\cdot\mathbf{x}}\textrm{ }\mathrm{F}^{-1}_{\mathbf{k''}\rightarrow\mathbf{x}}\bigg[e^{iz\sqrt{k^2-k_x^{''2}-k_y^{''2}}}\\
&\textrm{ } \mathrm{F}_{\mathbf{x'}\rightarrow\mathbf{k''}}\big[e^{i\mathbf{k}\cdot\mathbf{x'}}S(\mathbf{x'},\mathbf{k})\big]\bigg],
\label{fullProp}
\end{aligned}
\end{equation}
where F (F$^{-1}$) indicates a (inverse) Fourier transform. This equation is very similar to Fourier optics field propagation equations and can be easily evaluated with numerical Fourier methods. However, (\ref{fullProp}) works for any field regardless of coherence. Also note that operationally for a discrete phase-space, if we reshape it into a 2D distribution $(\mathbf{x},\mathbf{k})$, we are propagating each spatial frequency column separately. 

To propagate (\ref{genDiracS}), we can follow a similar procedure making the change $\rho\rightarrow U\rho U^\dagger$, such that
\begin{equation}
U =  \exp{(i f \frac{k_x^2+ k_y^2}{2 k})}\exp{(-i k \frac{x^2+ y^2}{2 f})}\exp{(i z \frac{k_x^2+ k_y^2}{2 k})}, 
\end{equation}
where $z$ is the distance from the object to the lens and $f$ is the lens focal length. Standard Fourier propagation methods can then be used to find the distribution at the object.

\subsection{Code Availability}
Code used for data processing and propagation is available upon request from the corresponding author.

\subsection{Data Availabilility}
The data that support the findings of this study are available from the corresponding
author upon reasonable request.

\section{Acknowledgements}
S.H.K. thanks G.A. Howland for insightful discussions. S.H.K, D.J.L, and J.C.H acknowledge support from the Air Force Office of Scientific Research Grant FA9550-16-1-0359, and from Northrop Grumman Grant 058264-002.  J.S. acknowledges support from the National Research Council Research Associate Programs, and funding from the OSD ARAP QSEP program. Any opinions, findings and conclusions or recommendations expressed in this material are those of the author(s) and do not necessarily reflect the views of AFRL.

\section{Author Contributions}
 S.H.K conceived the idea and performed the experiment. D.J.L. provided the algorithms for data analysis. J.S. provided theoretical support. J.C.H. supervised the project. S.H.K. prepared the manuscript with contributions from all other authors.
 
 \section{Competing Interests}
 The authors declare no competing financial interests.

\bibliography{DiracMatrix.bib}

\end{document}